\documentclass[12pt]{article}
\usepackage{color}
\usepackage{graphicx}
\RequirePackage[cdot,textstyle,amssymb]{SIunits}
\RequirePackage{units}
\RequirePackage{sistyle}
\SIdecimalsign{.}
\SIthousandsep{\,}
\SIproductsign{\times}
\SIunitsep{\,}
\SIunitspace{\cdot}
\SIunitdot{\cdot}
\usepackage{mathtools}
\usepackage{authblk}
\RequirePackage{soul}
\RequirePackage{journames}

\begin{document}

\title{Domain wall pinning in a circular cross-section wire with modulated diameter}

\author[1]{A. De Riz}
\author[2]{B. Trapp}
\author[3]{J. A. Fernandez-Roldan}
\author[2]{Ch. Thirion}
\author[2]{J.-Ch. Toussaint}
\author[1]{O. Fruchart}
\author[1]{D. Gusakova}
\affil[1]{Univ. Grenoble Alpes, CNRS, CEA, Spintec, F-38000, Grenoble, France}
\affil[2]{Univ. Grenoble Alpes, CNRS, Institut NEEL, F-38000, Grenoble}
\affil[3]{Institute of Materials Science of Madrid, CSIC, 28049, Madrid, Spain}

\maketitle \tableofcontents

\newpage

\section*{Abstract}

Domain wall propagation in cylindrical nanowires with modulations of diameter is a key phenomenon to design physics-oriented devices, or a disruptive three-dimensional magnetic memory. This chapter presents a combination of analytical modelling and micromagnetic simulations, with the aim to present a comprehensive panorama of the physics of pinning of domain walls at modulations, when moved under the stimulus of a magnetic field or a spin-polarized current. For the sake of considering simple physics, we consider diameters of a few tens of nanometers at most, and accordingly domain walls of transverse type. Modeling with suitable approximations provides simple scaling laws, while simulations are more accurate, refining the results and defining the range of validity of the models. While pinning increases with the relative change of diameter, a key feature is the much larger efficiency of pinning at an increase of diameter upon considering current rather than field, due to the drastic decrease of current density related to the increase of diameter.

\section{Introduction}
\label{Introduction}

\subsection{Fundamental and technological motivations for domain wall pinning}
\label{Domain wall pinning: technological issue}


The interest for domain walls in one-dimensional conduits is both for the sake of physics and for technological concepts. As regards physics, considering domain walls in nearly one-dimensional systems allows one to reduce the number of internal degree of freedom to a minimum. In the limit of cylindrical wires with a diameter typically below seven time the dipolar exchange length $l_\mathrm{ex}=\sqrt{2A_\mathrm{ex}/\mu_0 M_\mathrm{S}^2}$, with $A_\mathrm{ex}$ the exchange stiffness and  $M_\mathrm{S}$ spontaneous magnetization, one can neglect variations of magnetization across the wire section, boiling down the description of the domain wall to a one-dimensional problem~\cite{bib-THI2006}. In any case, compared with extended thin films this reduces the possible complexity of the wall, obviously easing the understanding of any phenomena related with domain-wall motion, \textsl{e.g.} precessional dynamics and spin-torques. As regards technology, domain walls have been proposed as means to store\cite{bib-BAR2006,bib-PAR2008,bib-LAV2013}, transport and process information\cite{bib-ALL2002,bib-ALL2005}.

It may be desirable to modulate the energy landscape of a domain-wall in such a one-dimensional conduit. This may include potential barriers or potential wells. On the fundamental side, such modulations can allow to repeatedly initialize the system with a domain wall at a precise location. This is especially useful to implement time-resolved measurements in a pump-probe scheme, which requires the averaging of reproducible events, including the preparation of a given type of domain wall\cite{bib-HAY2007}. Also, energy barriers may be used to confine a domain wall in a segment of finite length to ease its investigation\cite{bib-FRU2018b}. On the applied side, a digital memory device requires that bits of information are allocated a specific physical location. Thus, domain walls may be forced to remain in potential wells, or conversely, be separated by energy barriers. Among others, this prevents that successive walls in a conduit merge together, which would induce the loss of information. Also, similar to the argument given above for fundamental devices, defining a precise starting position can be helpful to clock circuits, for instance in the case of logic functions involving several domain walls.

The modulation of potential along has been largely developed and exploited in planar strips based on thin film and lithography technologies. Most are based on the modulation of geometry, which is easily achievable with lithography. This includes notches\cite{bib-YOK2000,bib-HAY2007}, protrusions\cite{bib-BRY2007} or more complex designs such as connection to other magnetic pads\cite{bib-LEW2010}. Other means have been demonstrated, such as stray field from neighboring magnetic pads\cite{bib-BEG2011} or domain walls\cite{bib-SAM2013}, ion irradiation\cite{bib-VOG2011,bib-SER2013} or reprogrammable electric-field gating\cite{bib-BAU2013}.

\subsection{Types of pinning for nanowires}
\label{Type of pinning: geometry, material}

In the present chapter we focus on cylindrical conduits, which we will call nanowires. Magnetic nanowires have been synthesized routinely for several decades, mostly by \textsl{e.g.} electroplating in polymer or anodized aluminum templates\cite{bib-FER1999a,bib-SOU2014,bib-FRU2018d}. This synthesis methods presents constraints to design modulations of the potential for domain walls, however also offers opportunities, with respect to flat strips. There exist essentially two designs, which have been developed experimentally and considered theoretically in the past ten years.

The first route for creating a potential landscape, is through the geometry of the wire, involving the longitudinal modulation of the diameter. Indeed, the energy of a domain wall sensitively depends on the wire (local) diameter, involving changes in both exchange and dipolar energy. The most commons means to achieve such a modulation are multistep anodization\cite{bib-SAL2018b,bib-FRU2018d} or pulsed anodization\cite{bib-LEE2010b} of aluminum. While the versatility is lower than with lithography for strips, a large variety of designs has been demonstrated. More exotic routes exist, such as pulsed plating followed by etching\cite{bib-CHA2012}, or the alternation of wire and tubes\cite{bib-SAL2013b,bib-NEU2013}. The focus of the present work is restricted to the diameter modulation of a plain wire.

The second route for creating a potential landscape, is through the longitudinal modulation of the material. While this is analogous to strips processed with local irradation or gating, it is more straightforward and versatile to achieve in nanowires, by changing the growth conditions during synthesis. The ways to achieve this are multibath anodization for more versatility, or pulsing the plating potential in a bath with several metal salts, for a faster implementation\cite{bib-DUB1999,bib-BOC2017}. Note that one may use various magnetic materials, especially varying the composition of compounds\cite{bib-FER2018}, or non-magnetic materials such as Cu\cite{bib-BOC2017,bib-BRA2017b}.

\subsection{Existing theories and experiments}
\label{Review-TheoryExperiments}

The one-dimensional landscape model for domain walls is probably one of the earliest problems tackled in magnetism to explain the physics of coercivity, as described by the Becker-Kondorski model\cite{bib-BEC1932,bib-BEC1939,bib-KON1937,bib-KON1940}. A key conclusion is that while domain walls are found at the bottom of energy wells at rest, the depinning field is associated with local maxima of slope of the potential, themselves coinciding with inflexion points of the potential curve. We will see that this concept is still applicable for the more specific theories developed in our contribution. Later on, the one-dimensional landscape model was use again in specific cases by A.~Aharoni and followers, again in the context of the physic of coercivity. Potential wells and steps\cite{bib-AHA1960}, slopes\cite{bib-ABR1960} and others, were introduced and described. These effective models have been made more specific to the geometry of a nanowire, highlighting the local slope

A number of micromagnetic simulations have been made, considering linear modulations\cite{bib-ALL2009}, sharp single modulations\cite{bib-SAL2013}, sharp constrictions\cite{bib-CHA2012}, smooth modulations of various length\cite{bib-FER2018}. However, often the processes of domain-wall nucleation at a wire's end and the process of going through the modulation are not studied separately, thus not well describing the latter. Besides, some detailed models of walls at modulations have been proposed\cite{bib-ALL2011}, however their complexity does not allow to shed a general picture on the phenomenon of pinning. Overall, the existing literature shows interesting features, however does not provide a comprehensive view. This lack has been driven the present work, to deriving simple analytical scaling laws, and compare the field-driven and current-driven cases.

Finally, note that experimental reports of the interaction of domain walls at modulations of diameter are still scarce and incomplete. Letting aside reports of magnetometry of large assemblies of wires still in a matrix, or experiments on single wires, however not separating the physics of nucleation from the one of going through the modulation, only a handful of reports exist of domain-walls in diameter-modulated single wires\cite{bib-BER2016}. These do not provide a comprehensive quantitative picture at present.

\begin{figure}[!h]
  \centering
  \includegraphics[width=\linewidth]{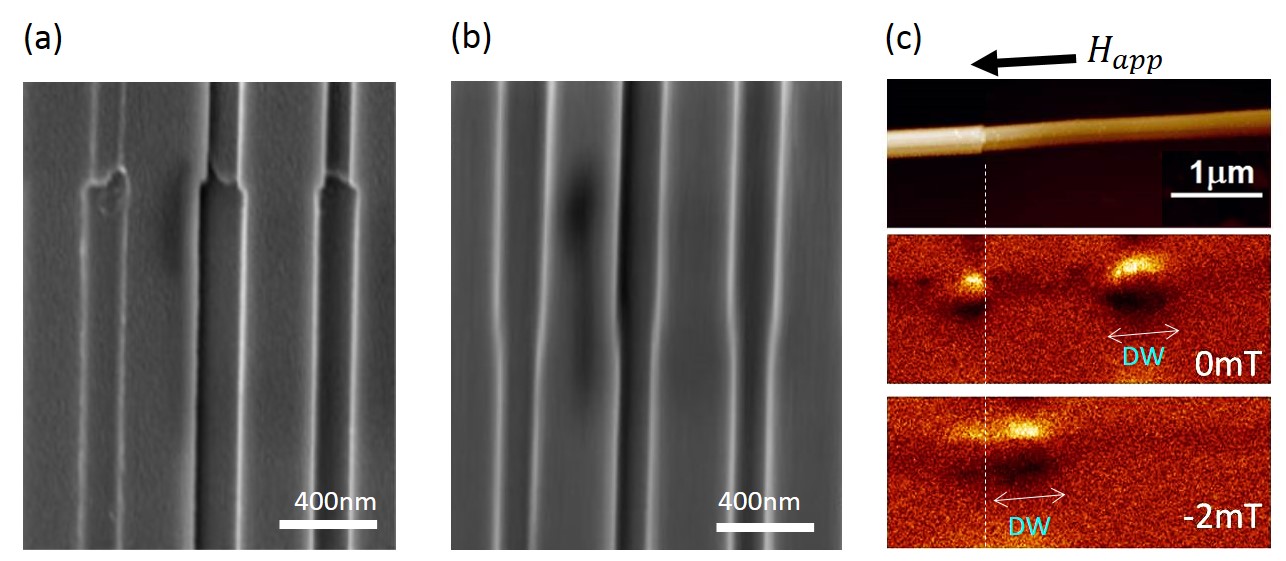}
  \caption{ (a) and (b) Scanning electron micrographs illustrating the existence of two different diameter transition
  geometries in the multisegmented aluminum oxide membranes from \cite{bib-FRU2018b} and \cite{TpappPhD}. (c) Topography of
  isolated multisegmented nanowire and magnetic force microscopy image showing the domain wall displacement after
  application of \textit{dc} field \cite{TpappPhD}.}
  \label{fig:MFM_image}
\end{figure}

\section{Theoretical background}
\label{Theoretical background}

The scope of the present section is to recall the basics of nanomagnetism in a circular cross-section nanowire comprising the domain wall, which are relevant for the concepts discussed in the following sections.  We start with the energy terms corresponding to ferromagnetic cylindrical nanowires with no modulation, which we will call \textsl{straight}. Then, we introduce the consequences of the diameter modulations, which imply the existence of an extra magnetic field related to the existence of the magnetic charges.

\subsection{Domain walls in cylindrical nanowires}
\label{Domain walls in cylindrical nanowires}
\begin{figure}[!h]
  \centering
  \includegraphics[width=\linewidth]{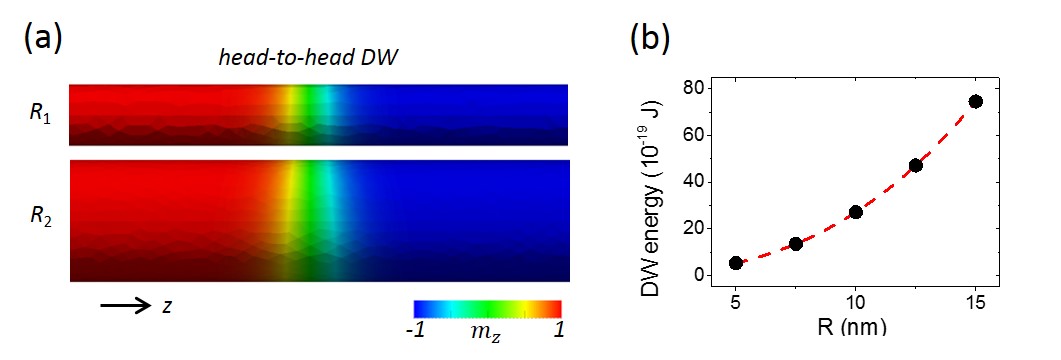}
  \caption{(a) Micromagnetic distribution of longitudinal magnetization for the transverse-like head-to-head domain wall for radius $R_1=\unit[5]{\nano\meter}$ and $R_2=\unit[10]{\nano\meter}$, obtained numerically using equation (\ref{eq:LLG}). (b) Simulated domain wall energy vs. diameter, ins straight wire diameter. The dashed curve corresponds is a third-order polynomial fit serving as a guide to the eye.}
  \label{fig:perfect_wire}
\end{figure}

Domain walls and domains in a ferromagnetic material are usually described within the framework of the micromagnetic theory. First introduced by W.F. Brown \cite{BROWN1940}, it is based on a continuous description of magnetization $\textbf{M}$ and of all other quantities. The norm of the magnetization vector is assumed to be constant and uniform, so that the local magnetization density can be written as a function of the lateral position $\textbf{r}$ and of time $t$ as $\textbf{M}(\textbf{r},t)=M_\mathrm{S}\, \textbf{m}(\textbf{r},t)$ with $M_\mathrm{S}$ being the spontaneous magnetization. A magnetization distribution can thus be described considering solely the unit vector $\textbf{m}(\textbf{r},t)$, which indicates the local orientation of the magnetization vector. Its time evolution is governed by the Landau-Lifshitz-Gilbert (LLG) equation
\begin{equation}
\frac{\partial \textbf{m}}{\partial t}=\gamma_0 \textbf{H}_\mathrm{eff}\times \textbf{m}+\alpha \textbf{m}\times \frac{\partial \textbf{m}}{\partial t}, \label{eq:LLG}
\end{equation}
where $\gamma_{0}>0$ is the gyromagnetic ratio and $\alpha$ is the phenomenological Gilbert damping factor. The effective magnetic field $\textbf{H}_\mathrm{eff}$ is defined as
\begin{equation}
\textbf{H}_\mathrm{eff}=-\frac{1}{\mu_0 M_\mathrm{S} V}\frac{\delta E}{\delta \textbf{m}} \label{eq:Heff}.
\end{equation}
It is related to the system's energy $E$, volume $V$ and magnetic permeability $\mu_0$. The first term of the LLG equation reflects the precession of the magnetization vector around the effective field, which may include the internal field contributions as well as the external applied field $\textbf{H}_\mathrm{app}$. Damping of this motion is described in the second term. This equation may be completed by the torque $\textbf{T}$ produced by the spin-polarized current. However, in this section we do not consider the phenomena related to the spin-polarized current, which are discussed in section~\ref{Modulation under applied current}.

In most cases of domain-wall motion, it is desirable to minimize extrinsic pinning such as due to spatial fluctuations of magnetic anisotropy, grain boundaries etc. Consequently, here we consider only magnetically-soft materials(such as~$\mathrm{Fe}_{20}\mathrm{Ni}_{80}$), in particular with no magnetocrystalline anisotropy. Thus, for a straight cylindrical wire\cite{hillebrans2006} under applied magnetic field, the energy $E$ of the system reads:

\begin{equation}
E=E_0+E_\mathrm{Z}, \label{eq:E}
\end{equation}
where
\begin{equation}
E_0=\frac{\mu_0}{2}\int_V \textbf{M}\cdot\textbf{H}_\mathrm{d}(\textbf{r})dV+A_\mathrm{ex}\int_V \left[ \nabla \textbf{m}(\textbf{r})\right]^2 dV, \label{eq:Energy0}
\end{equation}
\begin{equation}
E_\mathrm{Z}=-\mu_0 M_\mathrm{S} \int_V \textbf{m}(\textbf{r})\cdot \textbf{H}(\textbf{r})dV. \label{eq:Ez}
\end{equation}
$E_0$ is the internal energy. The first term in the equation (\ref{eq:Energy0})  corresponds to the magnetostatic contribution. The magnitude and orientation of the dipolar field $\textbf{H}_\mathrm{d}$ depend sensitively on the aspect ratio of the ferromagnet. The second term corresponds to the exchange contribution in its continuum form, with the exchange stiffness~$A_\mathrm{ex}$. When an external field $\textbf{H}_\mathrm{app}$ is applied, its contribution is described as the the Zeeman energy term (\ref{eq:Ez}).

For complex magnetic textures or non-trivial geometries, when simplifications of the LLG equation could not be done, the evolution of the magnetic system in time is solved numerically using appropriate micromagnetic codes. In this chapter all numerical simulations have been done using our home-built finite element freeware \textit{feeLLGood} (Finite Element Landau Lifshitz Gilbert equation Oriented Object Development)\cite{bib-ALO2014,bib-STU2015,bib-FEE}.
 The non-regular finite element mesh of \textit{feeLLGood} accurately describes the cylindrical geometry without creating an artificial numerical roughness at the cylinder surface. Moreover, \textit{feeLLGood}'s parallelized SCALFMM library\cite{bib-ScalFmm} based on so-called Fast Multipole Method for dipolar field calculation makes it competitive with usually less time consuming finite difference micromagnetic codes.

As we are considering soft magnetic material, the characteristic length scale of choice is the dipolar exchange length $l_\mathrm{ex}=\sqrt{2A_\mathrm{ex}/\mu_0 M_\mathrm{S}^2}$, resulting from the competition between exchange and magnetostatic energy contributions. Magnetization tends to be rather uniform over distances smaller than $l_\mathrm{ex}$, while it may rotate at larger scales under the influence of boundary conditions or dipolar energy. In cylindrical nanowires, depending on the wire diameter so far two different domain wall types have been theoretically predicted \cite{ivanov2013, hillebrans2006} and experimentally observed \cite{biziere2013, da_col2016, donnelly2017, stano2016}. Moderate wire diameters ($D<7 l_\mathrm{ex}$) considered in this chapter favor the formation of the so-called transverse-like domain wall \cite{FOR2002,SJamet2015}. In one dimension its profile along the $z$ axis is well described by $m_z=\tanh(z/\Delta)$ and $m_y=1/\cosh(z/\Delta)$, where $\Delta$ is the wall-width parameter. Fully tree-dimensional transverse-like domain wall distributions obtained by solving numerically equation (\ref{eq:LLG}) and corresponding energies are depicted in  figure \ref{fig:perfect_wire}.

\subsection{Geometry of modulation and potential barrier}
\label{Geometry of modulation and potential barrier}

\begin{figure}[!h]
  \centering
  \includegraphics[width=\linewidth]{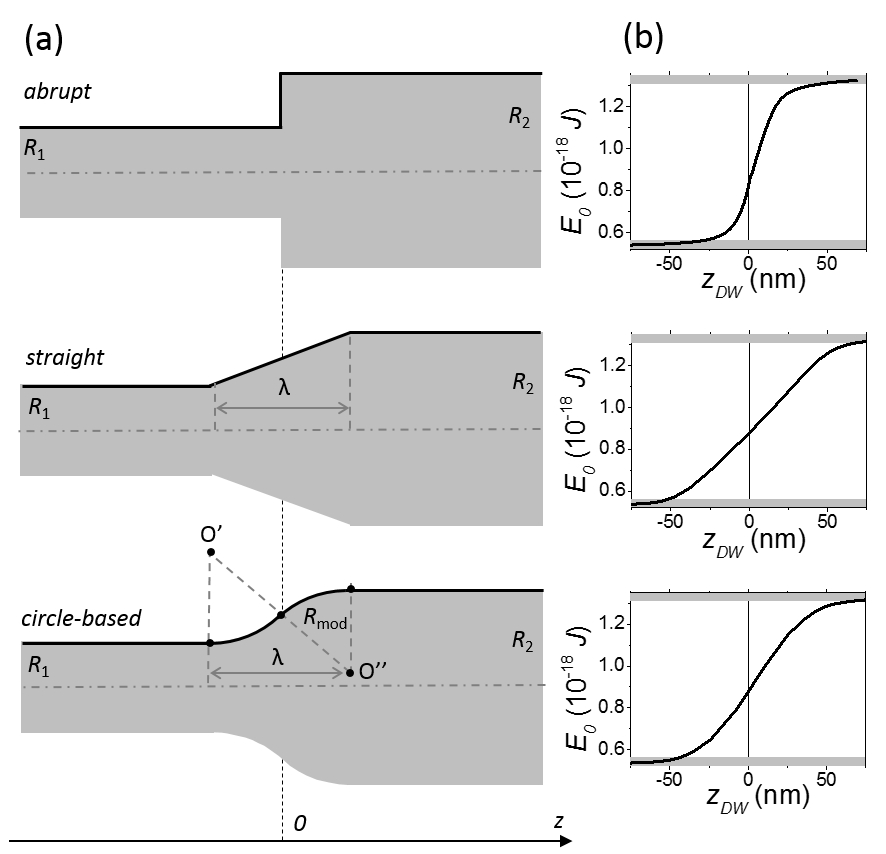}
  \caption{(a) Type of modulation geometry considered and (b) corresponding energy of the domain wall $E_0$ vs. its position $z_\mathrm{DW}$, from micromagnetic simulations. Parameters used for the energy plots are $R_1=\unit[5]{\nano\meter}$, $R_2=\unit[7.5]{\nano\meter}$, $\lambda=\unit[100]{\nano\meter}$ and $\mu_0 M_\mathrm{S}=\unit[5]{\tesla}$. Grey horizontal lines correspond to the energy of a straight wire with $R=\unit[5]{\nano\meter}$ and $R=\unit[7.5]{\nano\meter}$. }
  \label{fig:modulations_and_energies}
\end{figure}

A modulation of diameter induces a variation of the internal energy of the system, which depends on the longitudinal position of the domain wall. The center of the wall will be named~$z_\mathrm{DW}$, which does not mean that we are assuming a wall with zero thickness. In the following, we study four types of modulation profiles, which connect a smaller cross-section with radius $R_1$, to a larger cross-section with radius $R_2$: abrupt modulation, straight modulation, circular-based and tanh-based modulations, see figure \ref{fig:modulations_and_energies}(a).

The abrupt modulation is described by the simple step-function
\begin{equation}
R(z) = \left\{%
\begin{array}{ll}
R_1, &z<0,\\
R_2, &z>0.
\end{array}%
\right. \label{eq:abrupt_geometry}
\end{equation}

The straight modulation of with length $\lambda$ corresponds to the linear function
\begin{equation}
R(z) = \left\{%
\begin{array}{ll}
R_1,& z <-\lambda/2,\\
k z +s, & -\lambda/2<z<\lambda/2,\\
R_2, & z>\lambda/2,
\end{array}%
\right. \label{eq:straight_geometry}
\end{equation}
with $k=(R_2-R_1)/\lambda$ and $s=(R_2+R_1)/2$.

The circle-based profile allows for a smooth transition between smaller and larger cross-section parts
\begin{equation}
R(z) = \left\{%
\begin{array}{ll}
R_1,& z <-\lambda/2,\\
y_1-\sqrt{R_\mathrm{mod}^2-(z+\lambda/2)^2}, & -\lambda/2<z<0,\\
y_2+\sqrt{R_\mathrm{mod}^2-(z-\lambda/2)^2}, & 0<z<\lambda/2,\\
R_2, & z>\lambda/2,
\end{array}%
\right. \label{eq:circle-based_geometry}
\end{equation}
with $R_\mathrm{mod}=[(R_2-R_1)^2+\lambda^2]/[4(R_2-R_1)]$, $y_1=(R_2^2+2R_1R_2-3 R_1^2+\lambda^2)/[4(R_2-R_1)]$ and $y_2=(3 R_2^2-2R_1R_2-R_1^2-\lambda^2)/[4(R_2-R_1)]$. Is has been used in subsection \ref{Smooth modulation} and in section \ref{Modulation under applied current}  for the micromagnetic simulations. For the analytic calculations, the circle-based wire profile was approximately replaced by the tanh-based profile
\begin{equation}
R(z)=[R_1+R_2+(R_2-R_1)\tanh(4z/\lambda)]/2.
\label{eq:tanh-based_geometry}
\end{equation}
This is an analytic differentiable function, which approximates well the circle-based profile in the case of the gently sloping modulations studied in \ref{Smooth modulation} and \ref{Modulation under applied current}. For the gently sloping modulation with $(R_2-R_1 )<<\lambda$ the relative error made by tanh-based shape approximation instead of circular-based profile is less than 10 percents.

To illustrate the energy modification, in figure \ref{fig:modulations_and_energies}(b) we plotted the internal energy $E_0$ as a function of the position of the domain wall. These curves were obtained by solving the LLG equation (\ref{eq:LLG})
numerically for domain walls drifting freely from the broader part toward the thinner part of the wire in the absence of
any driving force. In that case we used $\alpha=1$, to approach a quasistatic situation. The energy of the domain wall is smaller in the thinner part of the wire. Handwavingly, this makes sense as the area and thus the volume of the domain wall, as well as its total magnetic charge and hence the dipolar energy, scale with the wire cross-section. Far from the modulation the value of $E_0$ recovers the value of energy of a
straight wire, as depicted by horizontal grey lines. The width of the transition between the lower and upper values of $E_0$, corresponds approximately to the modulations length~$\lambda$.

\subsection{Magnetic charges}
\label{Magnetic charges}

\begin{figure}[!h]
  \centering
  \includegraphics[width=8cm]{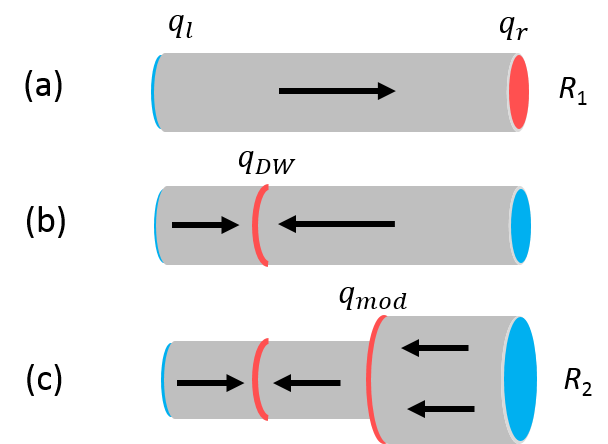}
  \caption{Schematics of magnetic charges distribution in (a) uniformly magnetized cylindrical wire, (b) cylindrical wire
  with head-to-head domain wall and (c) modulated diameter nanowire with head-to-head domain wall placed in the thinner
  part. Red color corresponds to the positive magnetic charge and blue one to the negative magnetic charge.}
  \label{fig:charge_conservation}
\end{figure}

By analogy with electrostatics based on Maxwell's equations, the magnetic volume and surface charges $\rho_\mathrm{m}=-M_\mathrm{S}\nabla \textbf{m}$ and $\sigma_\mathrm{m}=M_\mathrm{S}\left(\textbf{n}\cdot \textbf{m}\right)$, may be introduced as a source of the dipolar field $\textbf{H}_\mathrm{d}$ \cite{hubert_magnetic_1998,Coey2009}, where $\textbf{n}$ is the outward-pointing unit vector normal to the system surface. The expression for the dipolar field reads:
\begin{equation}
\label{eq:dipolarField}
\textbf{H}_\mathrm{d} (\textbf{r}) = \int \frac{\rho_\mathrm{m} (\textbf{r}') (\textbf{r}-\textbf{r}') }{4\pi |\textbf{r}-\textbf{r}'|^3} d^3\textbf{r}' + \oint  \frac{\sigma_\mathrm{m} (\textbf{r}') (\textbf{r}-\textbf{r}')}{4\pi |\textbf{r}-\textbf{r}'|^3} dS.
\end{equation}

\begin{figure}[!h]
  \centering
  \includegraphics[width=\linewidth]{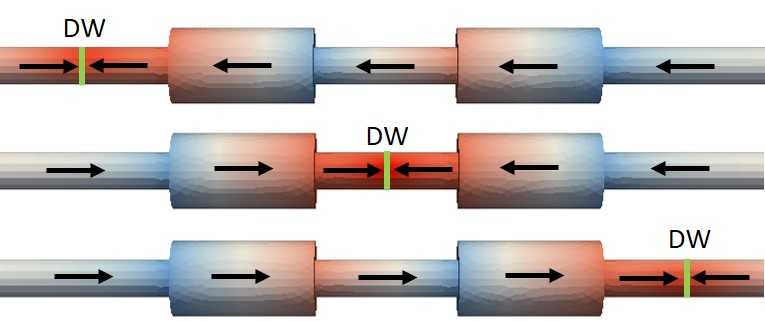}
  \caption{Magnetic potential $\phi_\mathrm{m}$ distribution for different positions of the head-to-head domain wall. Red~(resp. blue) color corresponds to positive~(resp. negative) values of $\phi_\mathrm{m}$. }
  \label{fig:magnetic_potential}
\end{figure}

In the case of the uniformly magnetized straight cylindrical wire (figure \ref{fig:charge_conservation}), each of the wire's end posses the total magnetic charge $q_1= \pm \pi M_\mathrm{S} R_1^2$. The total charge is an invariant, whose conservation imposes that the head-to-head domain wall in a cylindrical nanowire bears a total magnetic charge $q_\mathrm{DW}=2\pi M_\mathrm{S} R_1^2$. In the case of a modulation of diameter, the two end charges are different: $q_1= \pm \pi M_\mathrm{S} R_1^2$ and $q_2= \pm \pi M_\mathrm{S} R_2^2$. Considering for instance the case where the domain wall is clearly in the smaller-diameter part, $q_\mathrm{DW}=2\pi M_\mathrm{S} R_1^2$, and a charge is associated with the modulation: $q_\mathrm{mod}=\pi M_\mathrm{S} ( R_2^2-R_1^2)$ (figure~\ref{fig:charge_conservation}(c)). At this stage, we did not discussed which type of charge (volume or surface) contributes to the total charge. In all cases~(end, modulation, domain wall), the total charge is distributed both over volume charge density as well as the surface charge density, whose distributions are non-trivial. The micromagnetic distribution of the magnetic potential $\phi_\mathrm{m}$ related to the charge distribution ($\textbf{H}_\mathrm{d}=-\nabla \phi_\mathrm{m}$) illustrates this fact in figure~\ref{fig:magnetic_potential}. Most notably, the modulation charge $q_\mathrm{mod}$ gives rise to a magnetic dipolar field $H_\mathrm{mod}$, which we calculate in the next section~(subsection \ref{Magnetic field generated by the modulation}). We show that it tends to move the domain wall away from the modulation.

\subsection{Magnetic field generated by the modulation}
\label{Magnetic field generated by the modulation}

In this section we determine the magnitude and the direction of the magnetic field generated by the modulation of diameter. As regards its contribution of the energy of the system through its interaction with the domain wall, for simplicity we consider its value on the wire axis and at the center of the domain wall $z_\mathrm{DW}$.

Following Eq.\ref{eq:dipolarField}, the elementary magnetic field $d\textbf{H}$ generated by an element of magnetic charge $dq$ at the distance $r$ reads:
\begin{equation}
d\textbf{H}=\frac{dq}{4 \pi r^2}\frac{\textbf{r}}{r}.
\end{equation}
For the axisymmetrical charge distribution, which is the case here, the resulting magnetic field generated by the whole modulation, is aligned with the $z$ axis. While the the total charge of the modulation is fixed, its distribution over surface and volume contributions is not straightforward. Thus, some approximation that conserves the total charge of the modulation should be made. We assume that the magnetization vector in the modulation is strictly aligned with the $z$ axis at each point. This simplification limits the magnetic charge of the modulation to the surface charge $\sigma_\mathrm{m}$ only, while volume charges are zero. The surface charge approximation allows us to estimate the amplitude of the magnetic field generated by the modulation analytically in some specific cases.

\begin{figure}[!h]
  \centering
  \includegraphics[width=\linewidth]{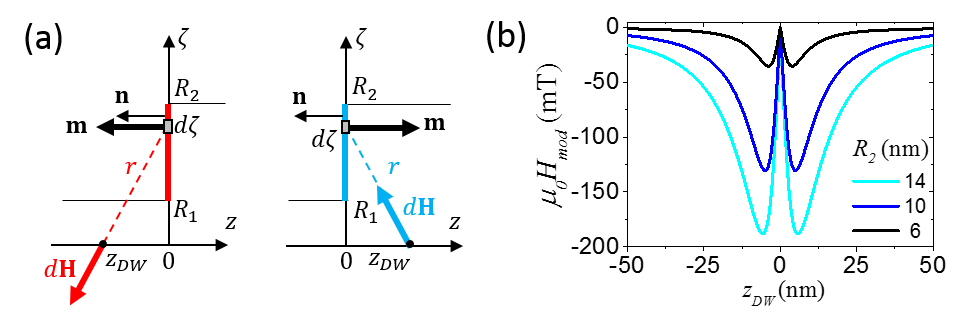}
  \caption{(a) Sketch of the magnetic field generated by the elements of the  magnetically charged axisymmetric surface in
  the presence of the head-to-head domain wall at the position $z_\mathrm{DW}$ for abrupt modulation. Red color corresponds to the positive surface charge and blue color to the negative one. (b) Magnetic field generated by the abrupt modulation $\mu_0
  H_\mathrm{mod}$ vs. domain wall position $z_\mathrm{DW}$ for several values of $R_2$. Parameters used for this plot are $R_1=\unit[5]{nm}$ and $\mu_0 M_\mathrm{S}=\unit[1]{T}$.}
  \label{fig:scheme_H_abrupt}
\end{figure}

\textit{Abrupt modulation}. Besides being close to applicable in some experimental cases, the abrupt modulation is a text-book case, from which the general features of the impact of a modulation on domain-wall motion can be easily illustrated.

With the above assumtions, the abrupt modulation is described by Eq.(\ref{eq:abrupt_geometry}). It is charged positively when the head-to-head domain wall is to the left of the modulation, and negatively when it is to the right of the modulation [Figure \ref{fig:scheme_H_abrupt}(a)]. The elementary magnetic field generated by the element $dS$ of the charged surface at the domain wall position $z_\mathrm{DW}$, and projected on the $z$ axis, reads:
\begin{equation}
dH_z(z_\mathrm{DW})=-\frac{\vert z_\mathrm{DW}\vert}{r}\cdot\frac{M_\mathrm{S} R dR}{2 r^2}.
\end{equation}
Here $dq=\sigma dS$, $\sigma=\pm M_S\left(\textbf{n}\cdot \textbf{m}\right)$, $dS=2 \pi R dR$, $\textbf{r}/r=\mp 1$, $dH_z =dH \vert z_\mathrm{DW}\vert/r$ and $r^2=z_\mathrm{DW}^2+R^2$. The summation of all contributions from element charges over the entire charged surface gives:
\begin{equation}
H_z(z_\mathrm{DW})=-\frac{M_\mathrm{S}}{2}\int_{R_1}^{R_2}\frac{\vert z_\mathrm{DW}\vert R dR}{\left(R^2+z_\mathrm{DW}^2\right)^{3/2}}.
\end{equation}
Upon integration, we obtain the magnetic field $H_{mod}\equiv H_z(z_{DW}) $ generated by the modulation at the center of the head-to-head domain wall
\begin{equation}
H_\mathrm{mod}(z_\mathrm{DW})=-\frac{M_\mathrm{S} \vert z_\mathrm{DW}\vert}{2}\left( \frac{1}{\sqrt{R_2^2+z_\mathrm{DW}^2}}-\frac{1}{\sqrt{R_1^2+z_\mathrm{DW}^2}}\right), \label{eq:Hmod_abrupt}
\end{equation}
which is plotted in Figure \ref{fig:scheme_H_abrupt}(b). The $H_\mathrm{mod}$ always opposes the head-to-head domain wall movement to the right, being negative for all $z_\mathrm{DW}$. In other words, the charges at the modulation tend to favor motion towards the part with smaller radius, similar to the energy of the domain wall itself.

\begin{figure}[!h]
  \centering
  \includegraphics[width=\linewidth]{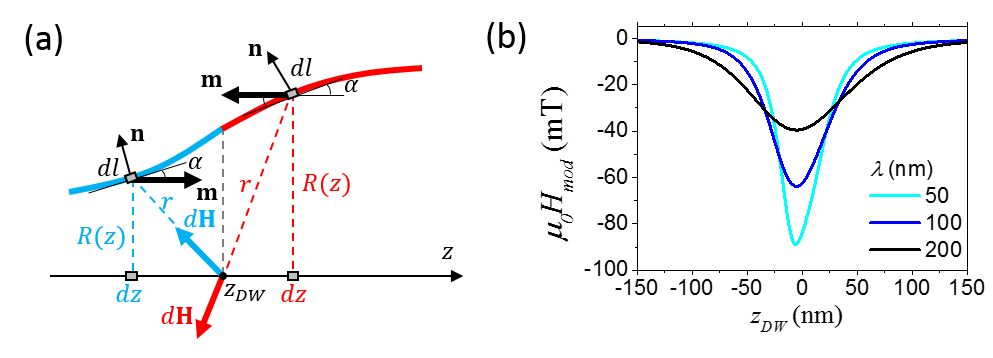}
  \caption{(a) Sketch of the magnetic field generated by the elements of the  magnetically charged axisymmetric surface in
  the presence of the head-to-head domain wall at the position $z_\mathrm{DW}$ for the modulation of arbitrary profile given by the   continuous function $R(z)$. Red color corresponds to the positive surface charge and blue color to the negative one. (b)~ Magnetic field generated by the tanh-based profile modulation of the length $\lambda$ vs. domain wall position $z_{DW}$ for several values of $\lambda$. Parameters used for this plot are $R_1=\unit[5]{nm}$, $R_2=\unit[10]{nm}$ and $\mu_0 M_\mathrm{S}=\unit[1]{T}$.}
  \label{fig:scheme_H_profile}
\end{figure}

\textit{Charged surface with arbitrary profile}. While the main physics is captured by the abrupt modulation, it is associated with an unphysical cusp of $H_\mathrm{mod}$ at the very center of the modulation. Besides, it may not be realistic for slowly-varying modulations such as found in some experimental cases. The present paragraph intends to describe such situations. Following the same method and assuming only surface charges, let us calculate the magnetic field generated by the modulation with an arbitrary profile given by the continuous function $R(z)$. As shown in Figure \ref{fig:scheme_H_profile}(a) the corresponding modulation surface is charged positively to the right of the head-to-head domain wall and negatively to the left of it. We may assume a stepwise jump of surface charges across the domain-wall, in the case of gentle modulations. The summation of all contributions from the entire charged surface reads:
\begin{equation}
H_\mathrm{mod}(z_\mathrm{DW})=-\frac{M_\mathrm{S}}{2}\int_{-\infty}^{\infty}\frac{\vert z_\mathrm{DW}-z\vert R(z)R'(z)dz}{\left[(z_\mathrm{DW}-z)^2+R^2(z)\right]^{3/2}}, \label{eq:Hmod_any_profile}
\end{equation}
This is derived from $\sigma_\mathrm{m}=\mp M_\mathrm{S} \sin(\alpha)$, $dS=2 \pi R(z)dl$, $dl=dz/\cos(\alpha)$, $\tan(\alpha)=R'(z)$ the derivative of~$R(z)$, $dH_z =dH \vert z_\mathrm{DW}-z\vert/r$ and $r^2=(z_\mathrm{DW}-z)^2+R(z)^2$.

Figure \ref{fig:scheme_H_profile}(b) depicts $H_\mathrm{mod}$ computed using Eq.\ref{eq:Hmod_any_profile} for a tanh-based profile given by the formula \ref{eq:tanh-based_geometry}. Similar to the case of abrupt modulation, $H_\mathrm {mod}$ opposes the head-to-head domain wall movement to the right. However, there is now no more cusp at $z_\mathrm{DW}=0$, and the maximum magnitude of $H_\mathrm{mod}$ is now found at the center of the modulation. Note that this maximum decreases sharply with increasing modulation length~$\lambda$.

\begin{figure}[!h]
  \centering
  \includegraphics[width=\linewidth]{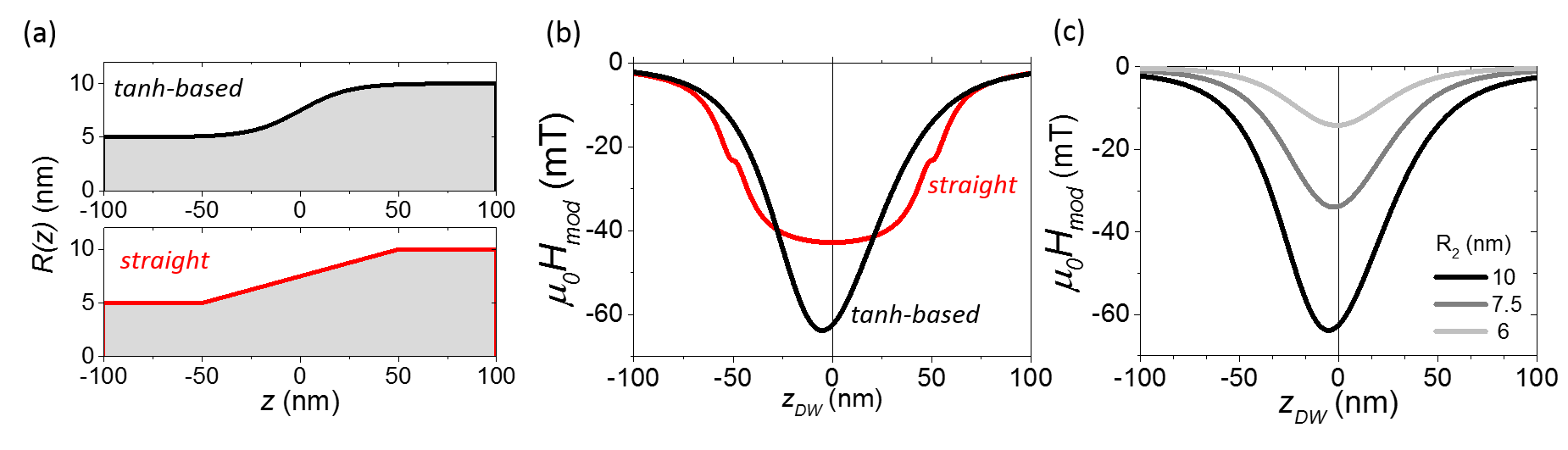}
  \caption{(a) Sketch of the tanh-based and straight profiles with $R_1=\unit[5]{nm}$, $R_2=\unit[10]{nm}$ and $\lambda=\unit[100]{nm}$. (b) Magnetic
  field generated by tanh-based and straight modulation profiles illustrated in (a). (c) Magnetic field generated by
  tanh-based modulation for several values of $R_2$. All curves are plotted for $\mu_0 M_\mathrm{S}=\unit[1]{T}$.}
  \label{fig:scheme_H_straight}
\end{figure}

\textit{Straight modulation}. In the case of the straight modulation of three segments [Figure \ref{fig:scheme_H_straight}(a)] given by the formula (\ref{eq:straight_geometry}), the integral in equation (\ref{eq:Hmod_any_profile}) should be calculated separately for each segment:
 \begin{equation}
H_\mathrm{mod}(z_\mathrm{DW}) = \left\{%
\begin{array}{ll}
\int_{-\lambda/2}^{\lambda/2}F(z_\mathrm{DW},z) dz, \\
-\int_{-\lambda/2}^{z_\mathrm{DW}}F(z_\mathrm{DW},z) dz+\int_{z_\mathrm{DW}}^{\lambda/2} F(z_\mathrm{DW},z) dz, \\
-\int_{-\lambda/2}^{\lambda/2}F(z_\mathrm{DW},z) dz ,
\end{array}%
\right.
\end{equation}
where
\begin{equation}
F(z_\mathrm{DW},z)=\frac{M_\mathrm{S}}{2}\frac{\left( z_\mathrm{DW}-z\right) (k z+s)k}{\left[(z_\mathrm{DW}-z)^2+(k z+s)^2\right]^{3/2}}.
\end{equation}
For very large $\lambda$ we may roughly estimate $H_\mathrm{mod}$ considering  $kz$ negligible in comparison to $s$. This gives the analytic expression for the field generated by the straight modulation in the following form

\begin{equation}
H_\mathrm{mod}(z_\mathrm{DW}) = \left\{%
\begin{array}{ll}
-\frac{ks M_\mathrm{S}}{2} \left( \frac{1}{\sqrt{(\lambda/2+z_\mathrm{DW})^2+s^2}}-\frac{1}{\sqrt{(\lambda/2-z_\mathrm{DW})^2+s^2}}\right),
\\
+\frac{ks M_\mathrm{S}}{2} \left( \frac{1}{\sqrt{(\lambda/2-z_\mathrm{DW})^2+s^2}}+\frac{1}{\sqrt{(\lambda/2+z_\mathrm{DW})^2+s^2}}-2\right),
\\
-\frac{ks M_\mathrm{S}}{2} \left( \frac{1}{\sqrt{(\lambda/2-z_\mathrm{DW})^2+s^2}}-\frac{1}{\sqrt{(\lambda/2+z_\mathrm{DW})^2+s^2}}\right).
\end{array}%
\right.
\end{equation}
Figure \ref{fig:scheme_H_straight} compares $H_\mathrm{mod}$ calculated for a straight modulation and a tanh-based profile.

\subsection{Energy of interaction}
\label{Energy of interaction}

In addition to the local terms [Eqs. (\ref{eq:Energy0}),(\ref{eq:Ez})] describing the domain wall behavior within the straight cylindrical wire, the energy of interaction of the domain wall with modulation charges $E_\mathrm{mod}$ must be considered. This is an extra contribution to the internal energy, while the domain wall moves through the modulation, however also at longer range. The total energy now reads:

\begin{equation}
E_\mathrm{tot}=E_0+E_\mathrm{Z}+E_\mathrm{mod}. \label{eq:EnergyTot}
\end{equation}
The derivative of energy with respect to the wall position, can be written under the form of an effective field. The one associated with the supplementary energy term $E_\mathrm{mod}$ is reads, for an axisymmetrical wire:

\begin{equation}
\frac{\partial E_\mathrm{mod}}{\partial z_\mathrm{DW}}=-\mu_0 q_\mathrm{DW} H_\mathrm{mod}(z_\mathrm{DW}). \label{eq:dEmod}
\end{equation}
It is unlikely that $E_\mathrm{mod}$ has an analytic expression in the case of an arbitrary modulation profile and arbitrary domain wall profile. In contrast, the field distribution $H_\mathrm{mod}(z_\mathrm{DW})$ can be derived analytically by making some assumptions, as shown in subsection \ref{Magnetic field generated by the modulation}. Besides, the $z$-derivative of energy may be sufficient, for example, to calculate the domain wall depinning field. In this case we do not need the energy $E_\mathrm{mod}$ expression but only its derivative, as the minimization of the total energy gives the domain wall pinned position.

\section{Modulation under applied magnetic field}
\label{Modulation under applied field}

\begin{figure}[!h]
  \centering
  \includegraphics[width=8cm]{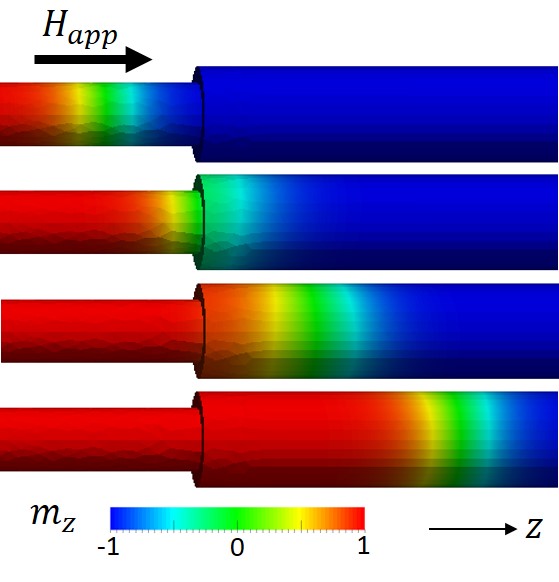}
  \caption{Head-to-head domain wall displacement under the applied magnetic field $H_\mathrm{app}$. The color scale bar indicates
  the longitudinal magnetization $m_z$. }
  \label{fig:under_H}
\end{figure}

In this section we focus on the case of domain wall behavior under a magnetic field applied along the wire’s axis [Figure \ref{fig:under_H}]. In particular we aim to calculate the critical field needed to depin the domain wall. As both the internal and the Zeeman energies are conservative, one may derive the critical field $H_\mathrm{crit}$ and corresponding critical domain wall position $z_\mathrm{crit}$ on the basis of the position-dependent domain wall energy.  In the majority of cases the purely analytical treatment of this problem is tricky or even impossible. For this reason, below we propose an analytical estimation of the $H_\mathrm{crit}$ in particular limit cases, which implies a number of simplifying hypothesis. Despite the limitations of the simplified approach, our analytical analysis focuses on the key ingredients and gives a very reasonable estimation of the behavior of the critical depinning field in response to the modulation parameters. The cases for which the assumptions used below are too drastic should be covered by micromagnetic simulations.

\subsection{Abrupt modulation}
\label{Abrupt modulation}

\begin{figure}[!h]
  \centering
  \includegraphics[width=\linewidth]{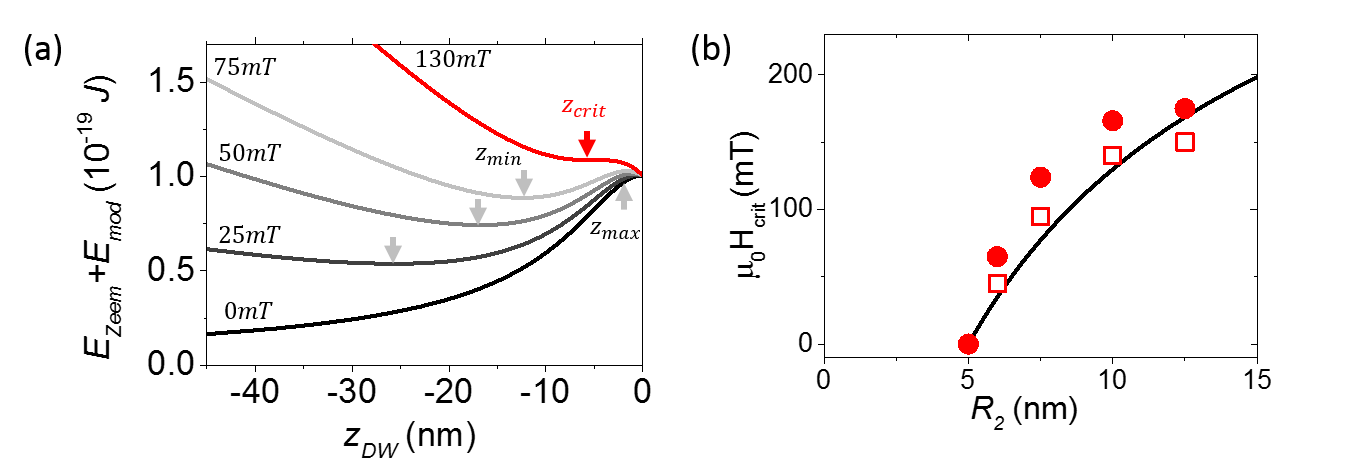}
  \caption{(a) Total energy $E_\mathrm{Z}+E_\mathrm{mod}$ versus domain wall position $z_\mathrm{DW}$ for several values of the applied field and $R_2=\unit[10]{nm}$. Vertical arrows show the pinned domain wall positions.  (b) Critical field value $H_\mathrm{crit}$ as a function of the larger radius $R_2$. Solid line corresponds to the analytic formula, solid circles and open squares correspond
  to micromagnetic simulations with $A_\mathrm{ex}=\unit[1\cdot 10^{-11}]{J/m}$ and reduced $A_\mathrm{ex}=\unit[0.25\cdot 10^{-11}]{J/m}$. All curves are plotted for $\mu_0 M_\mathrm{S}=\unit[1]{T}$ and $R_1=\unit[5]{nm}$.  }
  \label{fig:Hcrit_abrupt}
\end{figure}

In this subsection we estimate the critical applied field $H_\mathrm{crit}$ needed to depin the domain wall in a wire with an abrupt modulation of diameter, described by equation (\ref{eq:abrupt_geometry}) and visualized in figure \ref{fig:under_H}. The wire axis was taken as the $z$ direction. The modulation was centered at $z=0$ and $L$ is the total length of the wire. The head-to-head domain wall was prepared in the narrow section of the wire, and driven toward the larger section by applying a magnetic field.

Micromagnetic simulations suggest that for such modulation the transition between the two energy levels (or potential barrier) is relatively sharp (figure \ref{fig:modulations_and_energies}(b)). Moreover, magnetization is mostly perpendicular to the modulation surface, which gives the maximum surface charge $\sigma_\mathrm{m}=M_\mathrm{S} (\textbf{m}\cdot\textbf{n})$ and thus generates the large magnetic field of the modulation (equation (\ref{eq:Hmod_abrupt}) and figure \ref{fig:scheme_H_abrupt}(b)). In this case it is reasonable to assume that the key ingredient in domain wall pinning is the competition between applied magnetic field $H_\mathrm{app}$ and the magnetic field generated by the modulation $H_\mathrm{mod}$. Besides, the abrupt jump of diameter, and thus domain wall energy when crossing the modulation, makes that an abrupt jump may not describe all features of the total depinning process. Rather, it is illustrative to describe the long-range competition between the applied field contribution $E_\mathrm{Z}=-2 \mu_0 M_\mathrm{S} H_\mathrm{app}\pi R_1^2 z_\mathrm{DW}+\mathrm{Cste}$ and the energy of interaction between domain wall and modulation $E_\mathrm{mod}=-2\mu_0 M_\mathrm{S}\pi R_1^2 \int H_\mathrm{mod}(z_\mathrm{DW},z')dz'$. This explains the non-monotonic energy profile with domain wall position $z_\mathrm{DW}$, as shown in figure~\ref{fig:Hcrit_abrupt}(a). Note also, that we neglected the inner structure of the domain wall to derive the Zeeman energy, instead we considered the Zeeman energy of two adjacent uniformly magnetized domains on either sides of the domain wall’s center position, $z_\mathrm{DW}$.

The energy derivative $\partial (E_\mathrm{Z}+E_\mathrm{mod})/\partial z_\mathrm{DW}=0$ have extrema for $z_\mathrm{max}$ and $z_\mathrm{min}$. The the latter corresponds to the domain wall pinned position, while the former highlights the top of the energy barrier preventing further motion. Using equations (\ref{eq:Hmod_abrupt}), (\ref{eq:dEmod}) and applied field contribution, we obtain the expression which relates the applied magnetic field to the energy extrema:
\begin{equation}
H_\mathrm{app}=\frac{z_\mathrm{min,max} M_\mathrm{S}}{2}\left( \frac{1}{\sqrt{R_1^2+z_\mathrm{min,max}^2}}-\frac{1}{\sqrt{R_2^2+z_\mathrm{min,max}^2}} \right). \label{eq:Happ_vs_zmin_abrupt}
\end{equation}
At some critical value of applied field $H_\mathrm{crit}$ both extrema $z_\mathrm{min}$ and $z_\mathrm{max}$ converge into the same point~(an inflection point). $z_{crit}$ may be found using $\partial^2 (E_\mathrm{Z}+E_\mathrm{mod})/\partial z_\mathrm{DW}^2=0$. $z_\mathrm{crit}$ corresponds to the final  pinned position of the domain wall:
\begin{equation}
z_\mathrm{crit}=-\frac{R_1^{2/3}R_2^{2/3}}{\sqrt{R_1^{2/3}+R_2^{2/3}}}, \label{eq:Zcrit_abrupt}
\end{equation}
and the corresponding $H_\mathrm{crit}$ needed to depin the domain wall reads
\begin{equation}
H_\mathrm{crit}=\frac{M_\mathrm{S}}{2}\left( \frac{ z_\mathrm{crit}}{\sqrt{R_2^2+z_\mathrm{crit}^2}}-\frac{z_\mathrm{crit}}{\sqrt{R_1^2+z_\mathrm{crit}^2}}\right). \label{eq:Hcrit_abrupt}
\end{equation}
Figure \ref{fig:Hcrit_abrupt}(b) compares formula (\ref{eq:Hcrit_abrupt}) with micromagnetic simulation. This comparison reveals qualitatively and quantitatively similar tendencies. Note that simulations conducted with a value of $A_\mathrm{ex}$ reduced in comparison to that of the Permalloy-like material, fits slightly better the analytic results. Is may be explained by the more compact domain wall which probably better suits the model assumptions.

\subsection{Smooth modulation}
\label{Smooth modulation}

\begin{figure}[!h]
  \centering
  \includegraphics[width=\linewidth]{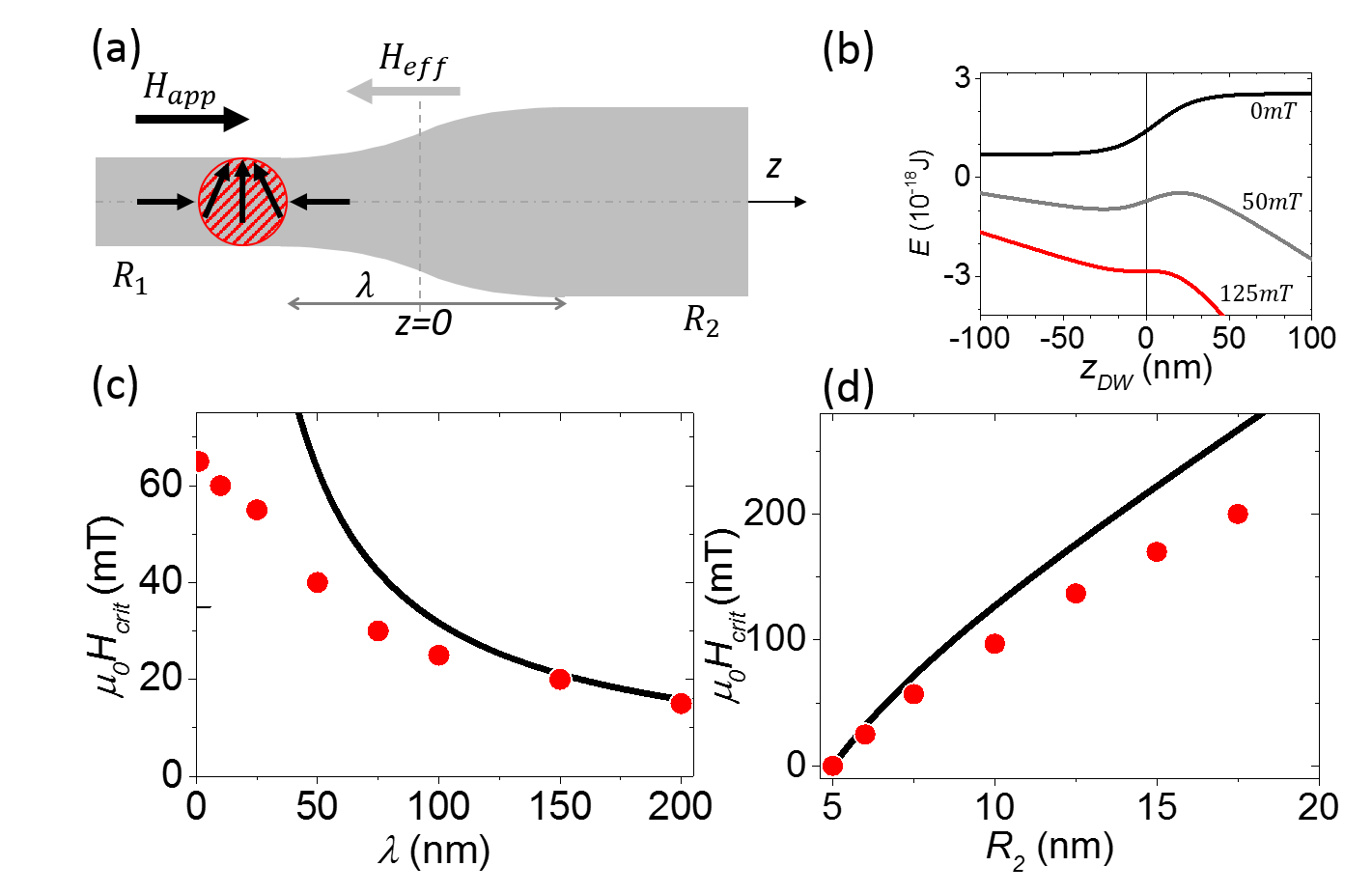}
  \caption{(a)Schematic illustration of the uniformly-charged sphere corresponding to the domain wall. (b) Domain wall
  energy $E_0+E_\mathrm{Z}$ as a function of the domain wall position, for several values of applied magnetic field. (c) Critical field $H_\mathrm{crit}$ as a function of modulation length $\lambda$ for $R_2=\unit[6]{nm}$. (d) Critical field $H_\mathrm{crit}$ as a function of larger radius $R_2$ for $\lambda=\unit[100]{nm}$. All curves are plotted for
  $\mu_0 M_\mathrm{S}=\unit[1]{T}$ and $R_1=\unit[5]{nm}$.}
  \label{fig:smooth_modulation_result}
\end{figure}

In this subsection we estimate the critical applied field $H_\mathrm{crit}$ needed to depin the domain wall in a smooth diameter modulation described by equation (\ref{eq:tanh-based_geometry}) and schematized in figure \ref{fig:smooth_modulation_result}(a). In practice, the modulation with length $\lambda$ was centered at $z=0$, and $L$ is the total length of the wire.  The head-to-head domain wall was prepared in the narrow section of the wire and was driven towards the larger section by applying a magnetic field. To determine the qualitative expression for $H_\mathrm{crit}$, we considered the domain wall and Zeeman energies $E_0$ and $E_\mathrm{Z}$, based on the magnetostatic, exchange and applied field contributions (Eqs.(\ref{eq:Energy0}),(\ref{eq:Ez})). For simplicity, here we omit the energy of interaction $E_\mathrm{mod}$ between the domain wall and the charges of the modulation. It has been shown in \cite{Fernandez2019} for smooth modulations, that the extension of the present model by including the $E_\mathrm{mod}$ does not have any qualitative impact, and results only in a slight shift in the total energy minima and maxima. Below, we introduce the approximations that can be used to estimate each energy term. The details of calculation may be found in \cite{Fernandez2019}.

For the dipolar energy, we considered that the magnetic charge $q_\mathrm{DW}=2M_\mathrm{S}\pi R^2$ \cite{kruger2011} carried by the head-to-head wall was uniformly distributed within the plain sphere of radius $R$, thus with a magnetic charge density $\rho_\mathrm{m}=3 q_\mathrm{DW}/4\pi R^3$. The real distribution of the magnetic charge is much more complex \cite{ferguson2015},\cite{hertel2015}. Nevertheless, our approximation leads to a compact analytical expression for the different energy terms and gives a reasonable order of magnitude. Note that this magnetic charge depends on the domain wall position $z_{DW}$, through $R(z_\mathrm{DW})$.

By analogy with electrostatics, a dipolar field $H_\mathrm{d}$ is generated by the charged plain sphere, with a total magnetostatic contribution $3\pi \mu_0 M_\mathrm{S}^2 R^3/5$. This contribution rapidly grows with the wire radius like $R^3$ which is consistent with the micromagnetic simulations of the domain wall energy plotted in figure \ref{fig:perfect_wire}(b) as a function of $R$. The exchange energy contribution can be estimated by applying the one-dimensional spin chain model \cite{hubert_magnetic_1998} with slowly varying magnetization. In this case $[\nabla \textbf{m}(r)]^2\approx(\pi/2R)^2$, so that the total exchange energy contribution equals $A_\mathrm{ex} \pi^3 R/3$. To estimate the Zeeman energy contribution, we neglected the inner structure of the domain wall and considered the Zeeman energy of two adjacent uniformly magnetized domains located at the domain wall’s center position, $z_\mathrm{DW}$. The domain wall energy excluding the integration constant then becomes:
\begin{equation}
E(z_\mathrm{DW})=\frac{3 \pi}{5}\mu_0 M_\mathrm{S}^2 R^3(z_\mathrm{DW})+\frac{A_\mathrm{ex}}{3}\pi^3 R(z_\mathrm{DW})-2 \mu_0 M_\mathrm{S} H_\mathrm{app}\pi \int_{-L/2}^z R^2(z)dz \label{eq:E_smooth}
\end{equation}
and is depicted in figure \ref{fig:smooth_modulation_result}(b). Note that is it compulsory analytically to consider the finite length of the wire, so that the Zeeman energy is finite.

Both local minima and local maxima are found using energy minimization $\partial E(z_\mathrm{DW})/\partial z_\mathrm{DW}$=0, which gives:

\begin{equation}
\frac{\partial R(z_\mathrm{DW})}{\partial z_\mathrm{DW}}\left(\frac{18}{5}+\frac{l^2_\mathrm{ex}\pi^2}{3R^2(z_\mathrm{DW})} \right)-\frac{4 H_\mathrm{app}}{M_\mathrm{S}}=0 \label{eq:dRdz_smooth}
\end{equation}
with $l_\mathrm{ex}=2 A_\mathrm{ex}/\mu_0 M_\mathrm{S}^2$. For a tanh-based profile and smooth modulation with $(R_2-R_1)/(R_2+R_1)<<1$, the coordinates of minimum and maximum of energy reads
\begin{equation}
z_\mathrm{max,min}=\pm \frac{\lambda}{4}\mathrm{arctanh}\sqrt{1-a H_\mathrm{app}}, \label{eq:zminmax_smooth}
\end{equation}
where $a=\frac{5\lambda}{9 M_\mathrm{S} (R_2-R_1)}\left[1+\frac{10 l_\mathrm{ex}^2 \pi^2}{27(R_1+R_2)^2}\right]^{-1}$. The coordinate of the energy minimum $z_\mathrm{min}$ corresponds to the domain wall pinned position. It corresponds to an internal effective field $H_\mathrm{eff}$ experienced at this point by the center of the domain wall:
\begin{equation}
H_\mathrm{eff}=-H_\mathrm{app}=-\left(1-\sqrt{1-H_\mathrm{app}}\right)/a. \label{eq:Heff_smooth}
\end{equation}
The domain wall depinning condition, at a given critical applied field value $H_\mathrm{crit}$, can be defined as the convergence of two energy extrema at the same point, $z_\mathrm{min}=z_\mathrm{max}$ (red curve in figure \ref{fig:smooth_modulation_result}(b)). Here we derive $z_\mathrm{crit}=0$ for $(R_2-R_1)/(R_2+R_1)<<1$ (the numerical solution of the equation (\ref{eq:dRdz_smooth}) without this assumption gives slightly different result \cite{Fernandez2019}). The corresponding critical field $H_\mathrm{crit}$ reads:
\begin{equation}
H_\mathrm{crit}=\frac{9 M_\mathrm{S} (R_2-R_1)}{5\lambda}\left(1+\frac{10 l_\mathrm{ex}^2 \pi^2}{27(R_1+R_2)^2}\right) \label{eq:Hcrit_smooth}
\end{equation}
and is depicted in figure \ref{fig:smooth_modulation_result}(c) and \ref{fig:smooth_modulation_result}(d) as a function of the modulation parameters. The domain wall repulsion from a modulation due to $H_\mathrm{mod}$, when not negligible, shifts $H_\mathrm{crit}$ towards higher values. Nevertheless, the analytical formula (\ref{eq:Hcrit_smooth}) provides a good estimation of $H_\mathrm{crit}$ and the relation between $H_\mathrm{crit}$ and geometric parameters. A key finding is that the critical field is  proportional to the slope of the modulation $(R_2-R_1 )/\lambda$, with a negligibly small exchange correction for small diameters.

The comparison between analytical formula (\ref{eq:Hcrit_smooth}) and micromagnetic simulations reveals qualitatively similar tendencies. Moreover, small $R_2/R_1$ ratios and long $\lambda$, corresponding to gently sloping modulations, ensure the best fit between the simulations and the analytical expression. The cases for which the assumptions used in this model are too drastic should be covered by micromagnetic simulations.

\subsection{Protrusion: double abrupt modulation}
\label{Protrusion: double abrupt modulation}

\begin{figure}[!h]
  \centering
  \includegraphics[width=\linewidth]{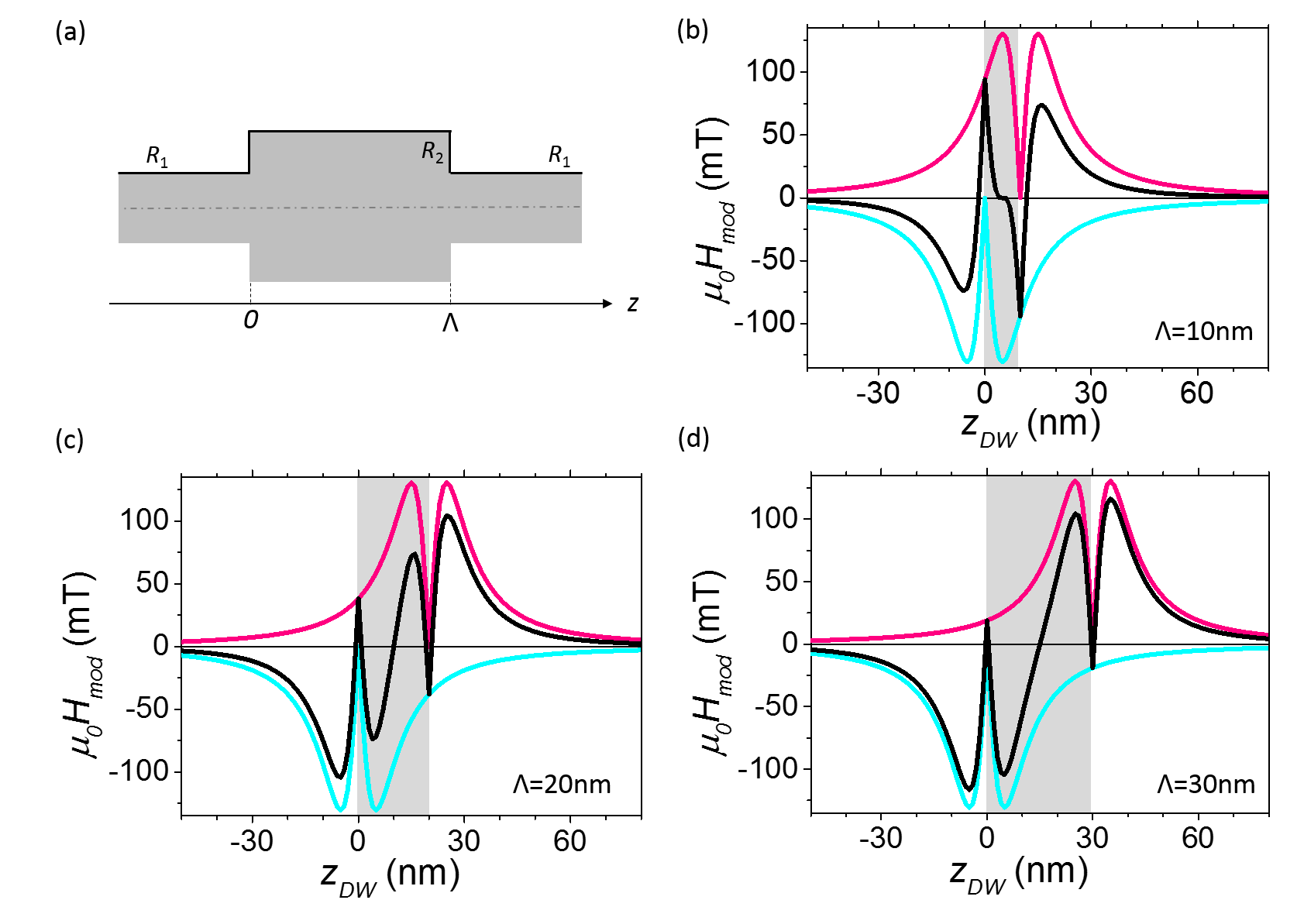}
  \caption{(a) Schematic of the protrusion geometry of the length $\Lambda$ and double abrupt radius modulation between smaller  one $R_1$ and  larger one $R_2$. (b),(c) and (d) Analytical curve for the magnetic field generated by a double modulation, for three values of the  protrusion length $\Lambda$. Red and blue lines correspond to the contribution of each modulation and black line to the   total resulting magnetic field $H_\mathrm{mod}$. All curves are plotted for $\mu_0 M_\mathrm{S}=\unit[1]{T}$, $R_1=\unit[5]{nm}$ and $R_2=\unit[10]{nm}$.}
  \label{fig:Hmod_double_abrupt}
\end{figure}

In this subsection we estimate the critical applied field $H_\mathrm{crit}$ needed to depin the domain wall in a wire with a protrusion with length $\Lambda$, as schematized in figure \ref{fig:Hmod_double_abrupt}(a) and given by the formula:
\begin{equation}
R(z) = \left\{%
\begin{array}{ll}
R_1,& z <0,\\
R_2, & 0<z<\Lambda,\\
R_1, & z>\Lambda.
\end{array}%
\right. \label{eq:protrusion_geometry}
\end{equation}
Similar to subsection \ref{Abrupt modulation}, the head-to-head domain wall was prepared in the narrow left section of the wire and driven towards the larger section by applying a magnetic field. To calculate $H_\mathrm{crit}$, the assumptions which may be done as well as the procedure to follow, are exactly the same. In order to calculate $H_\mathrm{crit}$ for any length $\Lambda$, we should examine the field $H_\mathrm{mod}$ created by the charges from both sides of the protrusion. Its expression reads
\begin{equation}
 \begin{split}
H_\mathrm{mod}=-\frac{M_\mathrm{S} \vert z_\mathrm{{DW}}\vert}{2}\left( \frac{1}{\sqrt{R_2^2+z_\mathrm{DW}^2}}-\frac{1}{\sqrt{R_1^2+z_\mathrm{DW}^2}}\right) \\
+\frac{M_\mathrm{S} \vert z_\mathrm{DW}-\Lambda\vert}{2}\left( \frac{1}{\sqrt{R_2^2+(z_\mathrm{DW}-\Lambda)^2}}-\frac{1}{\sqrt{R_1^2+(z_\mathrm{DW}-\Lambda)^2}}\right). \label{eq:Hmod_protrusion}
\end{split}
\end{equation}

\begin{figure}[!h]
  \centering
  \includegraphics[width=\linewidth]{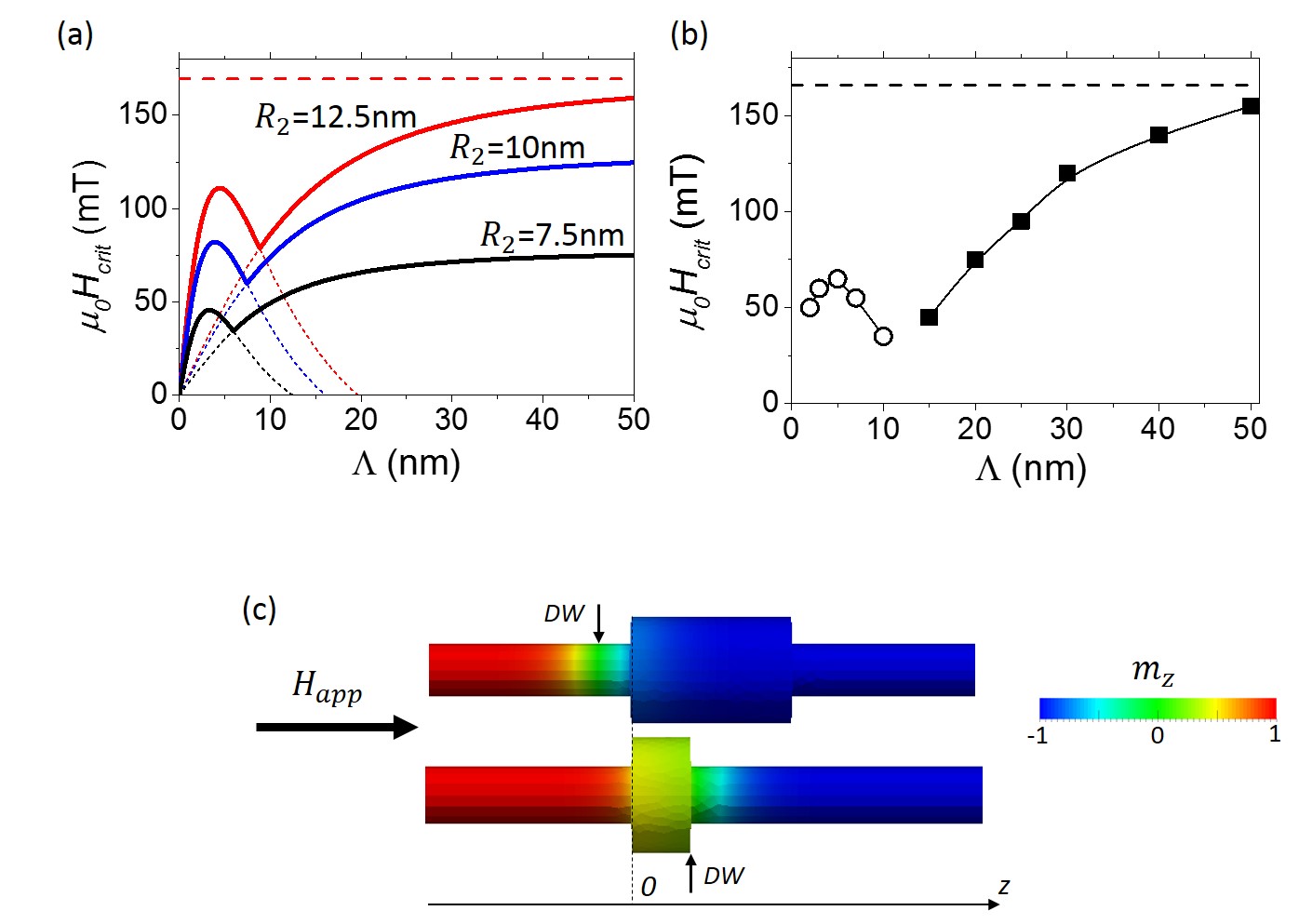}
  \caption{ (a) Critical field $H_\mathrm{crit}$ as a function of the protrusion length $\Lambda$ for several values of $R_2$. Red horizontal dashed line corresponds to the critical field value for $\Lambda=\infty$ (single abrupt modulation) and $R_2=\unit[12.5]{nm}$. (b) Micromagnetic simulation of the $H_\mathrm{crit}$ as a function of the protrusion length $\Lambda$ for $R_2=\unit[10]{nm}$. The branch with solid squares corresponds to the domain wall pinned to the left of the protrusion. The branch with open circles corresponds to the domain   wall pinned to the right of the protrusion. Black horizontal dashed line corresponds to the critical field value for $\Lambda=\infty$ (single abrupt modulation). (c) Domain wall pinned to the left of the protrusion for $\Lambda=\unit[30]{nm}$ and   domain wall pinned to the right of the protrusion for $\Lambda=\unit[10]{nm}$.  The color scale bar indicates the longitudinal   magnetization $m_z$. All graphs are plotted for $\mu_0 M_S=\unit[1]{T}$ and $R_1=\unit[5]{nm}$.}
  \label{fig:Hcrit_double}
\end{figure}

The magnitude of $H_\mathrm{mod}$ is plotted in figure \ref{fig:Hmod_double_abrupt}(b),(c) and (d) for different values of the protrusion length~$\Lambda$. Blue and red lines correspond to the contribution of each side of protrusion separately. The black solid line corresponds to the total $H_\mathrm{mod}$. Almost for all values of $\Lambda$ (except for very small $\Lambda$, i.e., for short modulation), $H_\mathrm{mod}$ has its deepest minimum to the left of the protrusion (for $z_\mathrm{DW}<0$). This means, the domain wall should be blocked to the left of the protrusion. For very small $\Lambda$, the situation maybe reversed (figure \ref{fig:Hmod_double_abrupt}(b)) and the domain wall may be pinned to the right of the protrusion. In order to quantify this phenomenon we followed the same procedure for the energy minimization similar to single abrupt modulation case and solved numerically $\partial^2 (E_\mathrm{Z}+E_\mathrm{mod})/\partial z_\mathrm{DW}^2$=0 condition. This numerical solution gives $H_\mathrm{crit}$ as a function of protrusion length $\Lambda$, and is plotted in figure \ref{fig:Hcrit_double}(a).

This gives rises to two regimes. As expected, for large $\Lambda$~(wide protrusion), the critical field $H_{crit}$ recovers the single abrupt modulation value. The pinning position is ahead of the protrusion, like for a single increase of diameter. Once the domain wall enters the protrusion, propagation proceeds beyond the center of the protrusion as the geometry favors expulsion, adding its effect to the Zeeman energy. For decreasing protrusion lengths the pinning field also decreases. This is related to the partial balance of the  repulsive charge on the left side of the protrusion and the attractive charge on the right side of the modulation. For very small values of $\Lambda$, the critical field $H_\mathrm{crit}$ has non-monotonic behavior, which correspond to the case with the domain wall pinned to the right of the modulation.

However, the model considers a domain wall with zero width, an hypothesis that may be put at stake for a short modulation. Besides, we stressed in section \ref{Abrupt modulation} that the model with abrupt modulation may not describe accurately the situation when the domain wall enters the modulation. For these reasons, it is important to check the situation with micromagnetic simulations. Surprisingly, these reproduce rather well the tendencies obtained in the simplified model, as shown in figures \ref{fig:Hmod_double_abrupt}(b) and \ref{fig:Hmod_double_abrupt}(c). Among others, the existence of two regimes is confirmed. This stresses that care needs to be taken in the analysis of experimental data, for which multiple pinning sites may not results from extrinsic imperfections.

\section{Modulation under applied current}
\label{Modulation under applied current}

\begin{figure}[!h]
  \centering
  \includegraphics[width=\linewidth]{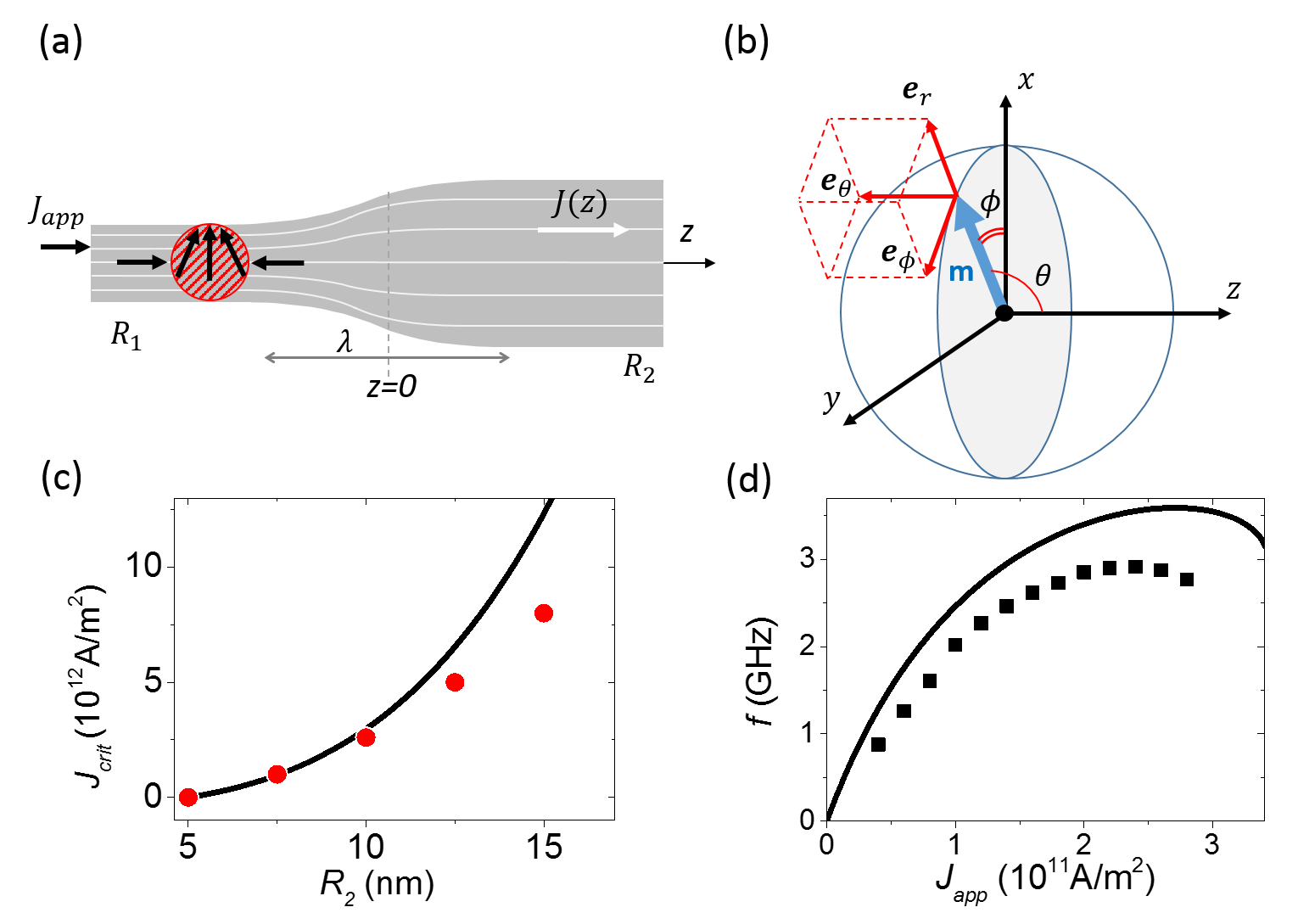}
  \caption{(a)Illustration of the domain wall under applied current in a modulated diameter wire. (b) Spherical coordinate
  basis $\lbrace\textbf{e}_r,\textbf{e}_\theta,\textbf{e}_\phi\rbrace$ for magnetization vector $\textbf{m}$ and Cartesian
  spatial coordinates $x$, $y$ and $z$. The magnetization vector is drawn in the particular position corresponding to
  $\theta=\pi/2$, so that $\textbf{e}_\theta=-\textbf{e}_z$.(c) Critical current density value $J_\mathrm{crit}$ versus the larger
  radius~$R_2$. The solid curve corresponds to the equation (\ref{eq:Jcrit_smooth}). Points correspond to the micromagnetic simulations.  (d) Domain wall rotation frequency $f$ in a pinned state vs. applied
  current density $J_\mathrm{app}$ for $R_2=\unit[10]{nm}$. The solid curve corresponds to the equation (\ref{eq:f_vs_Japp}). Points correspond
  to the micromagnetic simulations.  Curves (c) and (d) are plotted for $\mu_0 M_\mathrm{S}=\unit[1]{\tesla}$, $R_1=\unit[5]{nm}$, $\lambda=\unit[100]{nm}$, $\alpha=0.02$, $\beta=0.04$
  and $P=0.7$.}
  \label{fig:modulation_under_J_result}
\end{figure}

In this section we describe the domain wall behavior under applied current. For the sake of providing a realistic picture even when crossing the modulation, we immediately jump to the case of the smooth diameter modulation given by the tanh-based profile formula (\ref{eq:tanh-based_geometry}) and schematized in figure \ref{fig:modulation_under_J_result}(a).  The wire axis is again taken as the $z$ direction. The modulation of the length $\lambda$ is centered at $z=0$ and $L$ is the total length of the wire.  The head-to-head domain wall was prepared in the narrow section of the wire and driven towards the larger section by applying a spin-polarized current, with the electrons flowing from the narrow to the broad section.  Similar to subsection \ref{Smooth modulation}, we make several simplifying assumptions and focus on the key ingredients to estimate the critical current needed to depin the domain wall. We then use micromagnetic simulations to refine the analytic picture.

Under the applied spin-polarized current the domain wall motion obeys the LLG equation (\ref{eq:LLG}) generalized with the so-called adiabatic and non-adiabatic spin torques \cite{Thiaville2009}

\begin{equation}
\textbf{T}= -\frac{P \mu_\mathrm{B}}{e M_\mathrm{S}}\left[ \left(\textbf{J}\cdot \nabla\right)\textbf{m} -\beta \textbf{m}\times\left(\textbf{J}\cdot\nabla\right)\textbf{m}\right], \label{eq:Torque}
\end{equation}
with $P$ the spin-polarization ratio of the current, $\mu_\mathrm{B}$ the Bohr magneton, $e$ the (positive) elementary charge, $\beta$ the non-adiabatic coefficient and $\textbf{J}$ the electron current density. It is convenient to express the magnetization vector and the effective field in the spherical coordinates basis $\lbrace\textbf{e}_r,\textbf{e}_\theta,\textbf{e}_\phi\rbrace$ [figure \ref{fig:modulation_under_J_result}(b)] with $\textbf{m}=\textbf{e}_r$, $\textbf{H}_\mathrm{eff}=H_r \textbf{e}_r+H_\theta \textbf{e}_\theta+H_\phi \textbf{e}_\phi$ and $\dot{\textbf{m}}=\dot{\theta}\textbf{e}_\theta+\sin\theta \dot{\phi}\textbf{e}_\phi$. The circular symmetry of the nanowire leads to the energy rotational invariant, which implies $\partial/\partial\phi=0$ and thus $H_\phi=0$. Moreover, for simplicity we neglect any azimuth distortion of the domain wall, which corresponds to the 1-dimensional spin chain and implies $\nabla \phi=0$. We name $J_\mathrm{app}$ the current density in the narrow part of the wire, and assume that the electron current is parallel to the $z$ axis $\textbf{J}_\mathrm{app}=J_\mathrm{app}\textbf{e}_z$ (figure \ref{fig:modulation_under_J_result}(a)). This approximation is suitable for a smooth cross-section. Similarly, the electron current density is considered uniform in a cross-section and is related to the applied current density through $J_z(z)=R_1^2 J_\mathrm{app}/R^2(z)$.

After some algebra the LLG equation (\ref{eq:LLG}) augmented by the spin-torque (Eq.(\ref{eq:Torque})) takes the form

\begin{equation}
\dot{\theta}=\frac{\gamma_0}{1+\alpha^2}\alpha H_\theta-\frac{1+\alpha\beta}{1+\alpha^2}\frac{\sin\theta}{\Delta}\frac{P \mu_\mathrm{B}}{e M_\mathrm{S}}J_z(z), \label{eq:LLG_spherecoord_1}
\end{equation}
\begin{equation}
\sin\theta \dot{\phi}=-\frac{\gamma_0}{1+\alpha^2}H_\theta+\frac{\beta-\alpha}{1+\alpha^2}\frac{\sin\theta}{\Delta}\frac{P \mu_\mathrm{B}}{e M_\mathrm{S}}J_z(z), \label{eq:LLG_spherecoord_2}
\end{equation}
where $H_\theta=-(\mu_0 M_\mathrm{S} V)^{-1}\delta E/\delta\theta$, $\dot{\theta}=v \sin\theta/\Delta$, $v$ is the forward domain wall velocity and $\dot{\phi}=2\pi f$ is the angular domain wall velocity. Here we applied the useful property of a 1-dimensional domain wall profile $\partial\theta/\partial z=\sin\theta/\Delta$, where $\Delta$ is the wall-width parameter \cite{hillebrans2006,Malozemoff1979}.

For simplicity in the following we omit the internal domain wall structure and study the behavior of the magnetization vector in its center of the domain wall, where $\theta=\pi/2$. Let us assume that the domain walls settles at a given position~$z_\mathrm{DW}$, for a given value of applied current~$J_\mathrm{app}$. This corresponds to $\dot{\theta}=0$, which from Eq.(\ref{eq:LLG_spherecoord_1}) implies:

\begin{equation}
H_\theta=\frac{P \mu_\mathrm{B}}{e M_\mathrm{S}}\frac{(1+\alpha\beta)R_1^2J_\mathrm{{app}}}{\alpha \gamma_0 R^2(z)\Delta}. \label{eq:Htheta_vs_Japp}
\end{equation}

In the center of the domain wall, where $\textbf{e}_\theta=-\textbf{e}_z$, $H_\theta$ is parallel to the $z$ axis and is pointed to the negative $z$ direction, so that $H_\theta=-H_\mathrm{eff}$. From Eq.(\ref{eq:E}), $H_\theta=-H_0$ is the internal field, reflecting the $z$-dependence of the domain wall energy, which we calculated at any position in the field-driven model. From this, for a given $J_\mathrm{app}$ we can solve Eq.(\ref{eq:Htheta_vs_Japp}) to search for a $z$ value allowing an equilibrium position. Thus, by combining equations (\ref{eq:Heff_smooth}), (\ref{eq:Hcrit_smooth}) and  (\ref{eq:Htheta_vs_Japp}), we obtain a relation linking the applied current $(J_\mathrm{app})$ and the resulting steady-state position of the domain wall~$z_\mathrm{eq}$. Such a position exists for moderate current, however not for very large current. The cross-over determines the depinning current~$J_\mathrm{crit}$ in a smoothly-varying modulation:

\begin{equation}
J_\mathrm{crit}=\frac{9\alpha\gamma_0eM_\mathrm{S}^2}{20 P \mu_\mathrm{B}}\frac{(R_1+R_2)^3(R_2-R_1)}{R_1^2\lambda}\left(1+\frac{10 l_\mathrm{ex}^2 \pi^2}{27(R_1+R_2)^2}\right). \label{eq:Jcrit_smooth}
\end{equation}

Here, similar to subsection \ref{Smooth modulation}, we have assumed $z_\mathrm{crit}\cong 0$, and $\Delta \cong 2 R(z_\mathrm{crit})= R_1+R_2$. This law is plotted in figure~\ref{fig:modulation_under_J_result}. If we compare the domain wall behavior under applied field (figure \ref{fig:smooth_modulation_result}(d)) and under applied current (figure \ref{fig:modulation_under_J_result}(c)) in a smooth modulation, there is a major difference in the efficiency of the driving force in both cases. The critical field $H_\mathrm{crit}$ is almost linear with growing larger diameter whereas the growth of $J_\mathrm{crit}$ follows the power low of $R_2$. The reason is the decrease of local current density when the section broadens: not only does the domain wall energy increase, but the efficiency of spin torque decreases. Besides, $J_{crit}$ is proportional to the domain wall width-parameter which grows in the larger cross-section.

Figure \ref{fig:modulation_under_J_result}(c) compares the analytical solution with  micromagnetic simulations. The tendencies are similar, with even an excellent quantitative agreement in the limit of gentle modulation. This validates the model, and the above conclusion.

The model also predicts the frequency of precession  of the transverse component of magnetization of the wall, at the pinned position:
\begin{equation}
f=\frac{P\mu_\mathrm{B}}{2 \pi e M_\mathrm{S}}\frac{R_1^2 J_\mathrm{app}}{\alpha R^2(z)\Delta}. \label{eq:f_vs_Japp}
\end{equation}
The dominant effect is that of the internal field and not of the non-adiabatic spin torque, resulting in the $1/(\alpha\Delta)$ coefficient in this equation. This frequency is plotted in figure \ref{fig:modulation_under_J_result}(c), for which similar to subsection  \ref{Smooth modulation}, we estimated the wall-width parameter $\Delta$ as $\Delta\cong2R(z)$. Again, an excellent agreement is found with numerical simulation.

\section{Conclusion and perspective}

We have derived analytical models to describe how a magnetic domain wall may go through a modulation of diameter in a cylindrical nanowire, under the stimulus of either a magnetic field or a spin-polarized current. Scaling laws are derived, which may be used to quickly design a modulation to reach specific properties. While the depinning field scales with $M_\mathrm{S}$ and the slope of the modulation $(R_2-R_1)/\lambda$, the depinning current increases much faster with the geometrical strength of the modulation, due to the decrease of local current density, and the increase of wall width. The relevance of these laws are are confirmed by micromagnetic simulations, which reveal an excellent quantitative agreement for smoothly-varying modulations.

The quantitative investigation of experimental domain-wall pinning in modulations of diameter is still in its infancy. It has been carried-out under magnetic field so far, with scattered results, however pointing at the moderate strength of pinning compared with extrinsic pinning on material defects, when considering smoothly-varying modulations. The drastically higher efficiency of pinning under current, raises hope that modulations of diameter can be designed efficiently for spin-torque fundamental or applied devices.

\section*{Acknowledgements}
We acknowledge financial support from the French National Research Agency (ANR) (Grant No. JCJC MATEMAC-3D) and from the European Union’s Seventh Framework Programme (FP7/2007-2013) under grant agreement No.309589 (M3d). J. A. F.-R. is grateful for financial support from the Spanish MINECO under MAT2016-76824-C3-1-R and co-support from the ESF though BES-2014-068789 and EEBB-I-16-10934.

\newpage

\end{document}